\documentclass[aps,prl,twocolumn,showpacs,amsmath,amssymb,preprintnumbers,10pt,footinbib,superscriptaddress]{revtex4-1}
\usepackage{epsfig}
\usepackage{amsfonts}
\usepackage{amssymb}
\usepackage{enumitem}
\usepackage{amsthm}
\usepackage{subfig}
\usepackage{bm}
\usepackage{hyperref}
\usepackage{mathrsfs}
\usepackage{bbm}
\usepackage{cancel}
\usepackage[dvipsnames]{xcolor}
\usepackage{accents}
\usepackage{relsize}
\usepackage{float, slashed, graphicx, amssymb, amsmath}
\usepackage{shuffle}
\usepackage{tikz}
\usetikzlibrary{hobby,calc,shapes, positioning, backgrounds,trees, decorations, decorations.markings, decorations.pathmorphing, arrows, shapes.geometric, snakes}
\usepackage{tikz-feynman}
\tikzfeynmanset{compat=1.1.0}
\tikzfeynmanset{
graviton/.style={circle, draw=green!60, fill=green!5, very thick, minimum size=7mm}
}
\tikzfeynmanset{
codot/.style={/tikz/shape=circle,
/tikz/fill=red,/tikz/minimum size=0.1cm,/tikz/inner sep=1.8pt}
}
\tikzfeynmanset{sb/.style=
{/tikz/shape=circle,
/tikz/fill=black,
 /tikz/minimum size=0.3cm,/tikz/inner sep=1.pt
 } }
\tikzfeynmanset{myblob/.style=
{/tikz/shape=ellipse,
/tikz/fill=red,
 /tikz/minimum width=0.5cm,
 } }
 \tikzfeynmanset{myblobFF/.style=
{/tikz/shape=circle,
/tikz/fill=black,
 /tikz/minimum width=0.25cm,
 } }
 \tikzfeynmanset{myblob2/.style=
{/tikz/shape=rectangle,
/tikz/fill=red,
 /tikz/minimum width=0.1cm,/tikz/inner sep=1.8pt
 } }
  \tikzfeynmanset{myvertex/.style=
{/tikz/shape=circle,
/tikz/fill=black,
 /tikz/minimum width=0.05cm,
 } }
 \tikzfeynmanset{GR/.style={/tikz/shape=ellipse,/tikz/fill={rgb:black,1;white,2},/tikz/minimum size=0.3cm,} }
\tikzset{box/.pic={\filldraw[fill=black]  (0,0) circle (2.5pt);
				   \filldraw [fill=black] (0.5,0) circle (2.5pt);
			       \draw [line width=5pt] (0,0) -- (0.5,0);}}

\tikzset{wiggle/.style={decorate, decoration=snake}}

 \tikzfeynmanset{HV/.style=
{/tikz/shape=circle,
/tikz/fill={rgb:black,1;white,2},
 /tikz/minimum size=0.01cm,/tikz/inner sep=1.8pt
 } }

\usepackage[T1]{fontenc} % if needed

\makeatletter
\newcommand \UPlus {\mathop {\operator@font \uplus }\limits }
\makeatother
\makeatletter
\newcommand \Bigcup {\mathop {\operator@font \bigcup }\limits }
\makeatother
  \def\LabelNote#1{}%\smash{\hbox to\phipt{\raise1ex\hbox{\tiny[#1]}\hss}}}
 \def\Label#1{\label{#1}%
  \smash{\hbox to\phipt{\raise1ex\hbox{\tiny[#1]}\hss}}}
  
  \def\mdot{{\cdot}}

\def\veps{\varepsilon}

\def\nn{\nonumber}

\newcommand{\black}{\color{black}}

\def\spa#1.#2{\left\langle#1\,#2\right\rangle}
\def\spb#1.#2{\left[#1\,#2\right]}

\def\be{\begin{equation}}
\def\ee{\end{equation}}
\def\bea{\begin{eqnarray}}
\def\eea{\end{eqnarray}}  
\allowdisplaybreaks

\DeclareMathOperator{\arctanh}{arctanh}
\newcommand{\bz}{\mathbf{0}}
\newcommand{\bo}{\mathbf{1}}
\begin{document}

\preprint{
   SNUTP25-001
}

\title{Kerr Black Hole Dynamics from an Extended Polyakov Action}
\author{N. Emil J. Bjerrum-Bohr}
\email{bjbohr@nbi.dk}
\author{Gang Chen}
\email{gang.chen@nbi.ku.dk}
\affiliation{Center of Gravity, Niels Bohr Institute,\\ Blegdamsvej 17, DK-2100 Copenhagen, Denmark}
\affiliation{Niels Bohr International Academy, Niels Bohr Institute,\\ Blegdamsvej 17, DK-2100 Copenhagen, Denmark}
\author{Chenliang Su}
\email{chenliang.su@nbi.ku.dk}
\affiliation{Niels Bohr International Academy, Niels Bohr Institute,\\ Blegdamsvej 17, DK-2100 Copenhagen, Denmark}
\author{Tianheng Wang}
\email{tianhengwang@snu.ac.kr}
\affiliation{Center for Theoretical Physics, Seoul National University, \\
1 Gwana-ro, Gwanak-gu, 08826, Seoul, South Korea}

\begin{abstract} 
We examine a hypersurface model for the classical dynamics of spinning black holes. Under specific, rigid geometric constraints, it reveals an intriguing solution resembling expectations for the Kerr Black hole three-point amplitude. We explore various generalizations of this formalism and outline potential avenues for applying it to analyze the attraction of spinning black holes.
 \end{abstract}  


\maketitle

\section{Introduction}
Modeling spinning black holes, particularly those with significant classical spin, presents a formidable theoretical challenge deeply linked to the complex behavior of strongly coupled extended bodies. A prominent example of a spinning black hole is the Kerr black hole~\cite{Kerr:1963ud}, which, in the thorough analysis of Israel~\cite{Israel:1970kp}, was shown to be isomorphic to a hypersurface exhibiting a disc topology. It motivates the exploration of the hypersurface covariant action principle for spinning black holes, which we will pursue here.
 
In spinning black holes, the small-spin limit permits a perturbative framework utilizing the multipole expansion of an extended body. This approach is encapsulated in the well-established Mathisson-Papapetrou-Dixon equations~\cite{Mathisson:1937zz,papapetrou1951spinning,Dixon:1970zza,Dixon:1970zz,Dixon:1974xoz}, which have been extensively discussed in the literature. Recent advancements in scattering amplitudes~\cite{Vines:2017hyw,Arkani-Hamed:2017jhn,Guevara:2018wpp,Chung:2018kqs,Kosower:2018adc,Chen:2022kpm,Guevara:2019fsj,Arkani-Hamed:2019ymq,Aoude:2020onz,Chung:2020rrz,Guevara:2020xjx,Chen:2021kxt,Brandhuber:2021eyq,Kosmopoulos:2021zoq,Chiodaroli:2021eug,Bautista:2021wfy,Cangemi:2022bew,Ochirov_2022,Damgaard:2019lfh,Bern:2020buy,Vines:2018gqi,Maybee:2019jus,Haddad:2021znf,Bern:2022kto,Aoude:2022trd,Damgaard:2022jem,Bjerrum-Bohr:2023jau,Bianchi:2023lrg,Bautista:2023szu,Chen:2022clh,Bjerrum-Bohr:2023iey,DeAngelis:2023lvf,Brandhuber:2023hhl,Chen:2024mmm,Chen:2024bpf,Brandhuber:2024qdn,Chen:2024mlx,Cangemi:2023bpe,Bjerrum-Bohr:2024fbt,Bohnenblust:2024hkw,Bautista:2024emt,Bohnenblust:2024hkw,Akpinar:2025bkt,Brandhuber:2024bnz} and worldline formulations~\cite{Levi:2015msa,Mogull:2020sak,Jakobsen:2021zvh,Comberiati:2022ldk,Jakobsen:2021smu,Liu:2021zxr,Jakobsen:2022fcj,Menezes:2022tcs,FebresCordero:2022jts,Alessio:2022kwv,Scheopner:2023rzp,Ben-Shahar:2023djm,Kim:2023drc,Jakobsen:2023ndj,Gonzo:2024zxo,Haddad:2024ebn} , particularly pertaining to spinning point particles, have shown consistency with the multipole expansion for lower spin orders.  

Regardless, the situation becomes more complex for black holes exhibiting large spins---typically associated with superheavy masses in strong coupling regimes. Here, the limitations of the spin expansion become pronounced, and to achieve a reliable description applicable to both finite spin and finite gravitational coupling, one that exceeds the point-particle approximation while preserving a covariant coupling, it is necessary to elucidate an appropriate action principle. 

The analogy with the strong-coupling properties of matter in quantum chromodynamics is noteworthy in this context. It has yielded the Veneziano formula \cite{Veneziano:1968yb}, which extends hadronic interactions beyond the confines of perturbative quantum chromodynamics by employing concepts such as crossing symmetry and Regge behavior. The governing dynamics of such behavior are nowadays rephrased through the Nambu-Goto and Polyakov 1+1-dimensional string theory actions \cite{Brink:1976sc, Deser:1976rb, Polyakov:1981rd}, as well as higher-dimensional extensions, see {\it e.g.,}~\cite{Polchinski:1996na,Emparan:2009cs,Bjerrum-Bohr:2024wyw,Charalambous:2025ekl}.

Inspired by such ideas, this letter takes the first step toward exploring a new, comprehensive gravitational framework for characterizing strongly coupled spinning objects in a covariant way from an action principle. The model we propose is one with a rigid internal structure of a hypersurface with specific geometric constraints on the topology of \(\mathbb{S}^2 \times \mathbb{R}\) and dynamics governed by the equations of motion of a generalized Polyakov action. The metric and the Riemann tensor mediate the interaction between this hypersurface and the target spacetime. In this framework, the only additional assumption required is minimal couplings between fields that respect the fundamental symmetry of the system and the spin Supplemental condition. 

As we will see, this model effectively reproduces the Kerr three-point amplitude to arbitrary spin orders in a dimension-agnostic manner and, as a spinoff, accommodates a diverse array of effective generalized worldline theories. Thus, it opens numerous interesting new research avenues for characterizing spin dynamics, including explorations of the new action principle as a fundamental model for spin in gravitational systems that could provide new insights into higher multiplicity amplitudes, such as the Compton amplitude with large classical spin, which is needed for precision analysis of dynamics in binary merger events.

This letter is organized as follows: First, we explore hypersurfaces in flat and curved spacetimes. Next, we describe how the Kerr three-point amplitude at arbitrary spin orders can be situated within this analytical framework. Finally, we discuss several extensions of the formalism before concluding with our findings and highlighting interesting open questions for further exploration.

\section{Modelling the dynamics of a spinning black hole on a hypersurface}
Aiming to model a classical spinning black hole covariantly and at all spin orders while accounting for symmetry and physical requirements, we are inspired to consider an action principle based on a higher-dimensional extension of the Polyakov action integrated over the coordinates of a $(2+1)$ hypersurface. 
\begin{align}\label{def:Polyakov}
 C \int_{\mathbb{M}^2\times \mathbb{R}} d^3 \sigma \sqrt{\gamma} \Big(\partial_{a} Z^{\mu} \partial_{b} Z^{\nu} \gamma^{ab} \eta_{\mu\nu}\Big),
\end{align}
where $C$ is a constant with dimension and $\mathbb{M}^2$ denotes a compact closed Riemann surface, which together with $\mathbb{R}$ are parametrized by the coordinate $\sigma^a=(\tau,\vec{\sigma})$, while $Z^\mu(\sigma)$ denotes the spacetime coordinate of a given point in the worldvolume. Unless mentioned otherwise, we use the worldline convention for coordinates. This action will preserve general diffeomorphism invariance on the worldvolume and the target space, and we will employ a rigid worldvolume metric $\gamma^{ab}$ that is $\tau$-independent and respects the rotational symmetries expected for the spinning object,  {\it i.e.}, resembling a disc topology. The metric we work with has the following spherically symmetric generic form
\begin{align}
	ds^2=d\tau^2-a^2_w\big( d\theta^2+f(\theta) d\phi^2\big),
\end{align}
where $f(\theta)$ is an arbitrary function and $a_w$ is a constant with  `length' dimension; we note that the necessity of a non-trivial worldvolume metric is a characteristic feature of this higher-dimensional framework and has no direct counterpart in traditional worldline and string theories.  %

Since the description of the spinning black hole depends non-trivially on the worldvolume metric, we are led to fix $f(\theta)=\sin^2(\theta)$ using physical constraints. For instance, we insist that a suitable candidate metric must satisfy the free equation of motion following from eq. \eqref{def:Polyakov} and admit solutions that match weak-field expectations, including $\tau$-independence (see the \emph{Supplemental Material}~\cite{SuppMat} for more details). We arrive at a free action in the following form 
\begin{align}\label{def:PolyakovS2}
 \!\!\!\!  S_0&=-\frac{m}{ 8\pi a_w^2}\int_{\mathbb{S}^2\times \mathbb{R}} d^3\! \sigma \sqrt{\gamma} \Big(\partial_{a} Z^{\mu} \partial_{b} Z^{\nu} \gamma^{ab} \eta_{\mu\nu} +1\Big)\,,
\end{align}
where we denote the hypersurface $\mathbb S^2 \times \mathbb R$ and have included a mass term normalized consistently with the requirement of reparameterization invariance of $\tau$ so that we reach the point mass action in the limit $a_w\to 0$. The equation of motion of $Z^\mu(\sigma)$ is readily derived by varying the action \eqref{def:PolyakovS2},
\begin{align}
    \Box Z^\mu =\gamma^{ab}\Big(\partial_a \partial_b Z^\mu-\Gamma_{ab}^c(\partial_c Z^\mu)\Big)=0\, ,
\end{align}
where $\Gamma^c_{ab}$ denotes the nontrivial connection due to $\gamma_{ab}$, with nonvanishing components  
$\Gamma_{\theta\phi}^\phi=\Gamma_{\phi\theta}^\phi=\cot(\theta)$ and $\Gamma_{\phi\phi}^\theta=-\sin(\theta)\cos(\theta)$.

 The solution to the free equation of motion controls the classical trajectory of a given point on the hypersurface. The solution separates into a center-of-mass coordinate and a relative coordinate compared to the center of mass, {\it i.e.}~$Z_{\mathbf{0}}^\mu=X^\mu_\mathbf{0}+Y^\mu_\mathbf{0}$. Here $X^{\mu}_{\mathbf{0}}=x^\mu+v^\mu\tau$ gives the straight-line trajectory of the center of mass. For the relative coordinate, we denote it $Y^\mu_{\mathbf{0}}$ and write it in terms of spherical harmonics as  
\begin{align}\label{eq:ZSol}
Y^{\mu}_{\mathbf{0}} &=  \sum_{l>j\geq 0} Y_{l,j}^\mu\,,\nn\\
 Y_{l,j}^\mu &=a_w  \Big( c_{l,j} \,\beta_{\rm x}^{\mu}\cos\big({\sqrt{l(l+1)}\over a_w} \tau + j \phi\big) \\
 &\ \ \ \ \ \ +{c'_{l,j}}\,\beta_{\rm y}^{\mu}\sin\big({\sqrt{l(l+1)}\over a_w} \tau + j \phi\big) \Big)\mathcal{P}_{l}^j\big(\cos(\theta)\big),\nn
\end{align}
where $\mathcal{P}_{l}^j$  is the harmonic Legendre polynomial and $\beta_{\rm x},\beta_{\rm y}$ are unit vectors in flat spacetime. It is natural to impose $c_{l,j}=c'_{l,j}$ to keep the rotational symmetry.  

A feature of our model is that different modes $Y_{l,j}^\mu$ are associated with the physics of the spinning object we analyze. We will interpret these modes as maps from the worldvolume to solutions on a spinning disc in the target space of the radius $a_{t(l,j)}\equiv a_w c_{l,j}$, see fig.~\ref{fig:1}. For example, $Y_{1,1}^\mu$ is the single covering map to the top and bottom surface of the spinning disc, while $Y_{2,1}^\mu$ is a double covering while still consistent with the starting point that the source of a Kerr black hole is a disk~\cite{Israel:1970kp}. We interpret the finite size of the spinning disc in the target space as a result of the balance between rotation and the hypersurface inertial tension. (Although interesting, it is beyond the scope of this Letter to consider superpositions of multi-modes with different frequencies or wrapping numbers that clash with classical principles. Likewise, we observe that modes such as \( Y_{l,0}^\mu \) depend on only the variables \( (\tau,\theta) \), making them dimension-reducing. Given our expectation that the source of a Kerr-like black hole is inherently two-dimensional, we will not address such modes either.)\\[-30pt]
\begin{figure}[hbt]
  \centering
  \begin{tikzpicture}
    \node at (0.1,0) {\includegraphics[width=0.4\linewidth]{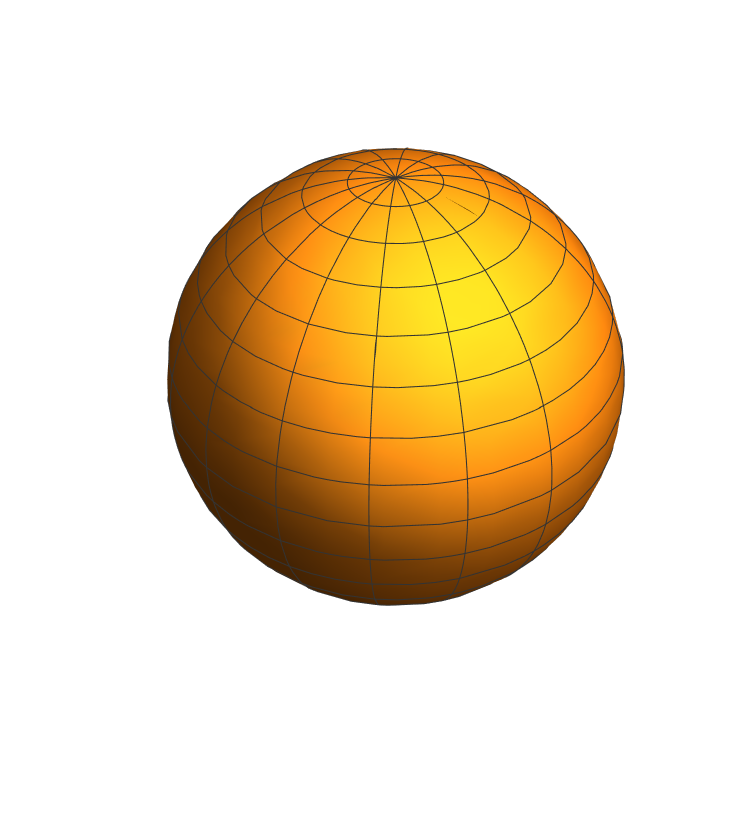}};
    \node at (2.5,0) {\raisebox{0.8\height}{$\xrightarrow{Y^{\mu}_{l,j}(\tau,\theta,\phi)}$}};
    \node at (5,0) {\includegraphics[width=0.43\linewidth]{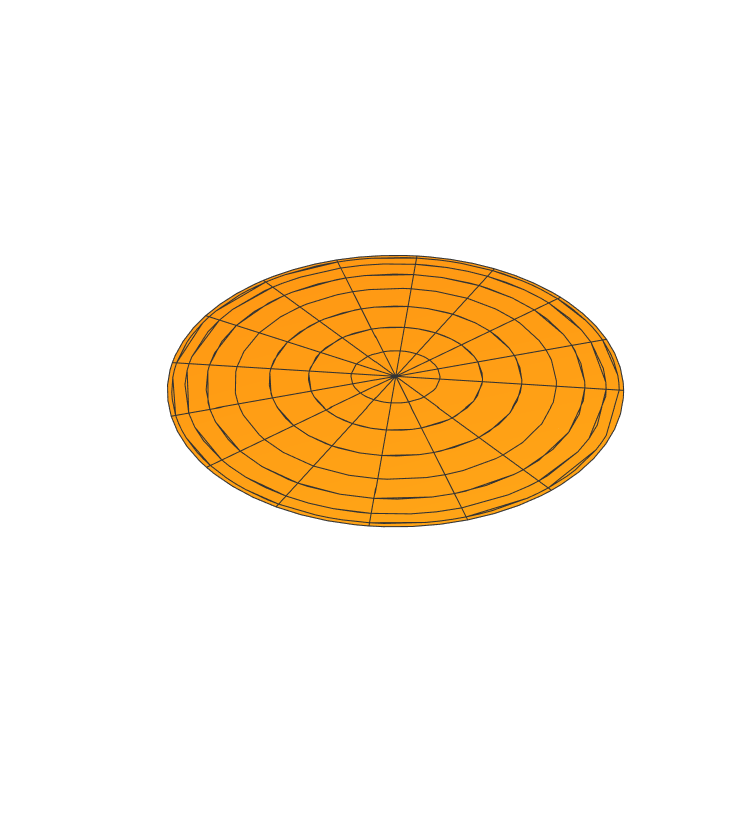}};
      \draw[dotted, ultra thick,blue] (5.1,0.12) -- (4.5,-0.5) node[anchor=north] {$a_{t(l,j)}$};
    \draw[-{Stealth[scale=1.]}] (5.1,0.12) -- (5.1,1.4) node[anchor=south] {\text{z}};
    \draw[-{Stealth[scale=1.]}] (5.1,0.12) -- (6.5,0.12) node[anchor=west] {\text{x}};
    \draw[-{Stealth[scale=1.]}] (5.1,0.12) -- (5.9,1.13) node[anchor=west] {\text{y}};
  \end{tikzpicture}\vskip-1.2cm

  \caption{\small Spinning hypersurface from worldvolume (left) to the disc in target spacetime (right), with coordinate axes labeled. The radius of the black hole is $a_{t(l,j)}$.}\label{fig:1}
\end{figure}
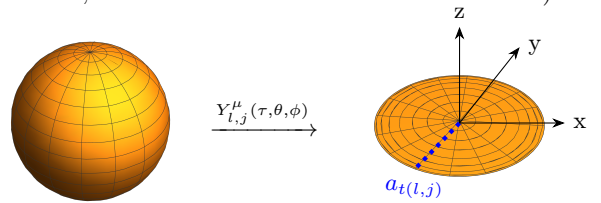\vskip-0.1cm

Using the solution for $Y^\mu_{\mathbf{0}}$, the angular momentum due to the rotation around the center of mass is 
\begin{align}\label{def:AngMom}
S^{\mu\nu}&={m a_w^2\over 4\pi a^2_w}\int_{\mathbb S^2}\,d\theta \,d\phi \,\sin(\theta)\,   (  Y^{\mu}_{l,j} \partial_\tau Y^{\nu}_{l,j}  -Y^{\nu}_{l,j}\partial_\tau Y^{\mu}_{l,j}   ) \nn\\
&= {\sqrt{l(l+1)}(l+j)!(m a_w)\over (2l+1)(l-j)!} c_{l,j}^2(\beta_{\rm x}^\mu \beta_{\rm y}^\nu-\beta_{\rm y}^\mu \beta_{\rm x}^\nu)\,.
\end{align}
Identifying \eqref{def:AngMom} with the standard spin tensor $S^{\mu\nu}=m\epsilon^{\mu\nu\rho\lambda}v_{\rho}a_\lambda$ (we impose the spin supplementary condition, $v\mdot \beta_{\rm x} = v\mdot \beta_{\rm y}= 0$ for orthogonal $\beta_{\rm x}$ and $\beta_{\rm y}$),
\begin{align}\label{eq:beta2a}
	S^{\mu\nu}&=m|a|(\beta_{\rm x}^\mu \beta_{\rm y}^\nu-\beta_{\rm y}^\mu \beta_{\rm x}^\nu), \nn\\
	v^{\mu}v^{\mu}-\eta^{\mu\nu}-{a^{\mu}a^{\nu}\over |a|^2}&= \beta_{\rm x}^\mu \beta_{\rm x}^\nu+\beta_{\rm y}^\mu \beta_{\rm y}^\nu \,.
\end{align}
We have spin length related to the radius of the $\mathbb{S}^2$ as follows
\begin{align}\label{eq:normC}
|a|={\sqrt{l(l+1)}(l+j)!\over (2l+1)(l-j)!} a_w c_{l,j}^2\,.
\end{align}

Extending this flat-space model to a generic curved background is straightforward. In principle, one can add any covariant coupling that respects the fundamental symmetries of the system. However, we have excluded certain types of terms due to physical reasons. For instance, to avoid introducing a potential term in the flat-space limit that spoils the free equation of motion, we exclude interactions of the form $[ (\mathcal D Z)^2]^n$ for $n\geqslant 2$, where $(\mathcal{D}Z)^2\equiv \partial_{a} Z^{\mu} \partial_{b} Z^{\nu} \gamma^{ab} \mathcal G_{\mu\nu}(Z)$ and  $\mathcal G_{\mu\nu}(Z)$  is the \emph{background} metric at the point $Z^\mu$. Explicit polynomial couplings in $Z^\mu$ are also excluded in favor of first-order derivative coupling of $Z^\mu$ to avoid coupling to the field $X^\mu$.

Our basic assumption is that each geometric point in $\mathbb{S}^2$ couples minimally to a background metric in a simple way, either directly or to the local Riemann tensor $\mathcal R_{\mu\nu\rho\lambda}(Z)$. (We do not consider higher-dimensional interactions, {\it i.e.}, higher derivatives of the metric or higher powers of the curvature. ) The  worldvolume geometry comes into play via the worldvolume metric $\gamma^{ab}$ and its respective Riemann tensor $\varrho^{abcd}$, with nonvanishing components $$\varrho^{\theta \phi \theta \phi}=-\varrho^{\phi\theta  \theta \phi}=-\varrho^{\theta \phi  \phi\theta}=\varrho^{\phi\theta   \phi\theta}={-1\over a_w^6\sin^2(\theta)}$$.

Thus, starting from the minimal action, taking into account the above considerations, we include the following additional couplings, which respect both the target space and the worldvolume diffeomorphism invariance
\begin{align}\label{eq:actionMS}
&S=-{m\over 8\pi a_w^2}\int_{\mathbb{S}^2\times \mathbb{R}} d^3 \sigma \sqrt{\gamma} \Bigg[(\mathcal{D}Z)^2 +1+ a_w^2\mathcal{R}_{\mu\nu\rho\lambda}(Z) \nn\\
    &\times\partial_a Z^\mu \partial_b Z^\rho \partial_c Z^\nu \partial_d Z^\lambda \Big( \gamma^{ab} \gamma^{cd}(\sum_{j=0}^\infty\xi_{2j+1} [(\mathcal{D}Z)^{2}]^j)\nn\\
    &
     +a_w^2 \varrho^{acbd}(\sum_{j=0}^\infty\xi_{2j+2} [(\mathcal{D}Z)^{2}]^j)\Big)\Bigg]\,.
\end{align}
To compute the classical three-point one-graviton amplitude using this action, we consider a perturbation around a flat metric 
$\mathcal{G}_{\mu\nu}(Z_{\mathbf{0}})=\eta_{\mu\nu}+\kappa\, h_{\mu\nu}(Z_{\mathbf{0}})$ with $\kappa=\sqrt{32\pi G_N}$ and 
identify the amplitude $A_3 (v,a,k)$ from the action using
\begin{align}
\begin{split}
   &\!\!\!\!\!\!\!\!\!\!\! \frac{ ie^{ik\mdot x}}{(2\pi)^4}2\pi\delta(mv\mdot k)A_3 (v,a,k) = -2h_{\mu\nu}(k)\frac{\delta S}{\delta h_{\mu\nu}(k)}\Big|_{h_{\mu\nu}\rightarrow 0}
    \end{split}\!\!\!\!\!\!\!\!\!\!\!\!
\end{align}
In order to evaluate this equation we rewrite $h_{\mu\nu}(Z_{\mathbf{0}})$ using its Fourier transform $h_{\mu\nu}(k)$
\begin{align}
	h_{\mu\nu}(Z_{\mathbf{0}})=\int {d^4k\over (2\pi)^4}\,h_{\mu\nu}(k) e^{ik\mdot X_{\mathbf{0}}(\tau)} e^{ik\mdot Y_{\mathbf{0}}(\sigma)}\,,
\end{align}
and take the outgoing on-shell graviton field to be a plane wave. (Without loss of generality, we will assume the graviton polarisation tensor is a product of two vector polarizations $\veps_\mu(k)\,\veps_\nu(k)$). Working to the leading classical post-Minkowskian order, it is helpful to define
\begin{align}\label{3pt amp from action}
    A^{(l,j)}_3(v,a,k)\equiv A_3(v,a,k)|_{Y^\mu_{\mathbf{0}}\rightarrow Y^\mu_{l,j}}\,,
\end{align}
and expand the various $e^{ik\mdot Y_{\mathbf{0}}(\sigma)}$ factors everywhere into powers of $(k\mdot Y_{\mathbf{0}}(\sigma))$. We note that it is straightforward to evaluate the worldvolume integral at a given order in $Y_{\mathbf{0}}(\sigma)$ and then perform the resummation after integration. 

Starting from the Polyakov part in eq.~(9), {\it i.e.}, ignoring all the Riemann tensor couplings, 
we illustrate the calculation of the $(1,1)$-mode three-point amplitude at the first post-Minkowskian order
\begin{align}
    & 2\pi i\, \delta(mv\mdot k)\,  A^{(1,1)}_{3,P}(v,a,k)=\frac{\kappa m}{4\pi}\int d\tau e^{ik\cdot v\tau}\int_{0}^{\pi}\sin(\theta)d\theta\nn \\
    &{\times}\int_{0}^{2\pi}d\phi\, \varepsilon_{\mu}\varepsilon_{\nu}(2v^\mu \partial_\tau Y_{1,1}^\nu+v^\mu v^\nu+\partial_{a} Y^{\mu}_{1,1} \partial^{a} Y^{\nu}_{1,1})e^{ik\mdot Y_{1,1}}\,.
\end{align}
Any term with an odd power of $Y^\mu_{1,1}$ in the $\phi$-integral is canceled out (including terms involving $\partial_a Y^\mu_{1,1}$). For even powers, the $\phi$-integral is readily evaluated. Each sequence-pair tensor structure can be rewritten using the identification provided by eq. \eqref{def:AngMom} and \eqref{eq:beta2a}. Schematically, we have the types of rewritings
\begin{align}
\begin{split}
    [\partial_\tau Y^{\mu}_{\mathbf{0}} Y^{\nu}_{\mathbf{0}}]&\rightarrow [S^{\mu\nu}]\\
 [Y^{\mu}_{\mathbf{0}} Y^{\nu}_{\mathbf{0}}],\ \ [\partial_b Y^{\mu}_{\mathbf{0}} \partial_b Y^{\nu}_{\mathbf{0}}] &\rightarrow \Big[v^{\mu}v^{\mu}-\eta^{\mu\nu}-{a^{\mu}a^{\nu}\over |a|^2}\Big]\, 
 \end{split},
\end{align}
where $b\in\{\tau,\theta,\phi\}$. We observe the presence of the desired kinematic variables. The $\phi$-integral value is independent of $\tau$; this renders the $\tau$-integral trivial and gives a factor $2\pi\delta(k\mdot v)$ as expected. For the contribution from $2v^\mu \partial_\tau Y_{1,1}^\nu$, only the odd power terms of $Y_{\mathbf{0}}$ in the series expansion of $e^{ik\mdot Y_{\mathbf{0}}}$ are nonvanishing. Using eq.~(8), the total contribution from these terms is $ i(v\mdot \veps)(k\mdot S\mdot \veps) \mathcal{I}_{P,o}^{(1,1)}$. Carrying out the $\theta$-integral, we arrive at 
\begin{align}
	\!\mathcal{I}_{P,o}^{(1,1)}
    &=\frac{3  |a|^2   \cosh \left(\frac{a_wc_{1,1} k\mdot a}{|a|}\right)}{(a_wc_{1,1})^2 (k\mdot a)^2}-\frac{3  |a|^3   \sinh \left(\frac{a_wc_{1,1} k\mdot a}{|a|}\right)}{(a_wc_{1,1})^3 (k\mdot a)^3}.
\end{align}
Considering a radius of the disk in the target space $a_w c_{1,1}$ as a Kerr black hole $|a|$ (recall $c_{1,1}=\frac{3}{2\sqrt{2}}$ from eq.~\eqref{eq:normC}), we arrive at functions such as $\cosh(k\mdot a)$ and $\sinh(k\mdot a)$. 
Similarly, the terms with the common factor $v^\mu v^\nu+\partial_{a} Y^{\mu}_{1,1} \partial_{b} Y^{\nu}_{1,1} \gamma^{ab}$ contribute only to the even orders in spin. The same procedure yields 
\begin{align}
\begin{split}
	\mathcal{I}_{P,e}^{(1,1)}  
	&=\frac{ \left(17 \left(a\mdot k\right){}^2+27\right) \sinh \left(a\mdot k\right)-27 a\mdot k \cosh \left(a\mdot k\right)}{8 \left(a\mdot k\right){}^3}
    \end{split}\, .
\end{align}
It follows that the amplitude from the Polyakov part is
\begin{align}
\begin{split}
    A^{(1,1)}_{3,P}(v,a,k)&=-i\kappa (mv\mdot \veps)( mv\mdot \veps\,  \mathcal{I}_{P,e}^{(1,1)} +i  k\mdot S\mdot \veps \, \mathcal{I}_{P,o}^{(1,1)} )
\end{split}\, .
\end{align}
We see that the minimal Polyakov action term completely determines $A_3^{(l,j)}$ up to ${\mathcal O}(|a|)$. It follows from the fact that interactions involving $\mathcal{R}_{\mu\nu\rho\sigma}$ contribute only at order $\mathcal{O}(|a|^2)$ and higher, as can be seen from spin counting: $a_w, Y^\mu_{l,j}$ each carry spin power one, while  $\partial_a Z^\mu_{\mathbf{0}}$ gives no extra spin power. 
These contributions from Riemann tensor couplings can be evaluated analogously and also lead to $\cosh(k\mdot a)$ and $\sinh(k\mdot a)$ for the $Y^\mu_{1,1}$ mode in the amplitude. We note that higher-order modes, which are not analyzed in detail here, give rise to distinct entire functions, such as generalized hypergeometric functions (see the \emph{Supplemental Material}~\cite{SuppMat}).  

We find that \eqref{eq:actionMS} evaluated for the $(1,1)$-mode suffices to give the three-point amplitude of any such black hole at the leading post-Minkowskian order.
To see this, we write the amplitude following from the generic action in a spin-expanded form up to $\mathcal{O}(|a|^s)$ and find that any amplitude admitting a regular spin expansion of only $(k\mdot a)$ can be reproduced by a unique set of $\xi_{1,2,\dots,s}$. We have verified this up to $s=99$. 
 
 We only list the final results for the first few terms in \eqref{eq:actionMS}, for example, {\it i.e.} $\xi_{1,2,4}$ to be arbitrary and all other interaction terms vanishing. The amplitude organised by a basis of entire functions, then is 
 \begin{align}\label{eq:A311}
	A_{3}^{(1,1)}(v,a,k)&=(-i\kappa)\sum_{i=1}^5c_{i}B_i, 
\end{align}
where $c_1={25\over 8}\xi_1+{9\over 4}\xi_2-{9\over 32}\xi_4$, $c_2={17\over 8}-{79\over 8}\xi_1-{9\over 4} \xi_2+{369\over 16}\xi_4$, $c_3=\frac{27}{32} (24 \xi_1-81 \xi_4-4)$, $c_4={3}\xi_1$,  $c_5=3-9\xi_1$ and basis elements $B_i$ are listed in tab.~\ref{tab:1}. 
\begin{table*}
    \centering
    \begin{tabular}{c|c|c|c|c|c}
        &$B_1=m^2 (v\mdot\veps)^2{\cal C}$ & $B_2=m^2(v\mdot\veps)^2{\cal S}$ & $B_3=m^2(v\mdot\veps)^2{\partial_{k\mdot a}\over k\mdot a}{\cal S}$ & $B_4=i(mv\mdot\veps)(k\mdot S\mdot \veps){\cal S}$& $B_5=i(mv\mdot\veps)(k\mdot S\mdot \veps){\partial_{k\mdot a}\over k\mdot a}{\cal S}$  \\ [5pt]\hline  
      $\chi$&  $\displaystyle\frac{(2 y^2-1)|b|^2 }{|b|^2-|a|^2}$ & $(2 y^2-1) {\cal A}_b$ & $\displaystyle ( y^2-{1\over 2}) ( {|b|^2\over |a|^2}(1-{\cal A}_b)+ {\cal A}_b)$ & $\displaystyle\frac{2y\sqrt{y^2-1}|a||b| }{|b|^2-|a|^2}$& $\displaystyle2y\sqrt{y^2-1}({\cal A}_b-1)\frac{|b|}{|a|}$\\ [5pt]\hline
     $ds^2_{(1)}$ &$\displaystyle{2 r\over r^2+|a|^2 \cos^2(\theta)}$&$\displaystyle{2\over |a|} {\cal A}$&$\displaystyle{(1-3 \cos^2(\theta )) W(r)\over 2|a|}- {\sin ^2(\theta ){\cal A}\over |a|} $&$\displaystyle { 4|a| r\sin^2(\theta)\over r^2+|a|^2 \cos^2(\theta)}$& $2\sin ^2(\theta )W(r)$ \nn\\
     &$[ds^2_{\rm flat}-2dt^2]$&$[ds^2_{\rm flat}-2dt^2]$&$[ds^2_{\rm flat}-2dt^2]$&$[d\phi dt]$& $[d\phi dt]$
    \end{tabular}
    \caption{We define $\displaystyle{\cal A}\equiv\arctan\Big({|a|\over r}\Big)$, $\displaystyle{\cal A}_b\equiv {|b|\over |a|}\arctanh\Big(\frac{|a|}{|b|}\Big)$, $\displaystyle W(r)\equiv \Big(1+{r^2\over |a|^2}\Big) {\cal A}- {r\over |a|}$, $\displaystyle{\cal S}\equiv{\sinh(k\mdot a)\over k\mdot a}$, $\displaystyle{\cal C}\equiv\cosh(k\mdot a)$. In the bending angle and metric, we omit the overall factor $\displaystyle{\kappa ^2 \sqrt{s}}/({16 \pi |b| \left(y^2-1\right)})$ and $G_N m$ respectively, where $s= m_1^2+m_2^2+2m_1 m_2 y$ and $y=v_1\mdot v_2$. We also have used the definition:  $ds^2_{\rm flat}\equiv\Big[ dt^2-{  \left(|a|^2 \cos ^2(\theta )+r^2\right)}/({ \left(|a|^2+r^2\right)})dr^2 - \left(|a|^2 \cos ^2(\theta )+r^2\right)d\theta^2- \left(|a|^2+r^2\right)  \sin ^2(\theta )d\phi^2\Big]$ .} 
    \label{tab:1}
\end{table*}
  The $\xi_2, \xi_4$ terms only contribute to even powers in spin due to the presence of $\varrho^{abcd}$.
Interestingly, we have discovered that we can recover a new representation of the Kerr black hole three-point amplitude (following from an extended Polyakov action with minimal coupling to the background) by fixing the coupling constants $\xi_i$ to be $\xi_1={1\over 3}$, $\xi_2=-{1\over 3^4}$, $\xi_4={4\over 3^4}$, $\xi_i=0$ for $i=3$ or $i>4$. Thus, we get,
\begin{align}\label{eq:actionMSKerr}
\begin{split}
S_{\rm Kerr}&=-{m\over  8\pi a_w^2}\int_{\mathbb{S}_2\times \mathbb{R}} d^3 \sigma \sqrt{\gamma} \Bigg[(\mathcal{D}Z)^2 +1\black\\
    &+ a_w^2\mathcal{R}_{\mu\nu\rho\lambda}(Z)\partial_a Z^\mu \partial_b Z^\rho \partial_c Z^\nu \partial_d Z^\lambda \\
    &\times\Big( {1\over 3}\gamma^{ab} \gamma^{cd}
     -{1\over 3^4}a_w^2 \varrho^{acbd}+{4\over 3^4}a_w^2 \varrho^{acbd}(\mathcal{D}Z)^{2}\Big)\Bigg]
\end{split}\, ,
\end{align}
which amazingly lands us directly on the three-point Kerr amplitude, see refs. \cite{Arkani-Hamed:2017jhn,Vines:2017hyw,Guevara:2018wpp,Chung:2018kqs,Chen:2022kpm}, 
\begin{align}
        &A^{(1,1)}_{3,\rm Kerr}(v,a,k) =\begin{tikzpicture}[baseline={([yshift=-1.0ex]current bounding box.center)}]
    \tikzstyle{every node}=[font=\small]	
    \begin{feynman}
         \vertex (a) {\(v, a\)};
         \vertex [right=1.cm of a] (f2) [dot]{};
         \vertex [right=.7cm of f2] (c){};
         \vertex [above=.8cm of f2] (d){};
         \diagram* {
         (a) -- [fermion,thick] (f2)-- [fermion,thick] (c),
         (f2)--[photon,ultra thick,momentum'=\(k\)](d)
         };
    \end{feynman}  
    \end{tikzpicture}=-i\kappa (B_1+B_4)\nn\\
       &=-i\kappa(mv\mdot \veps)
        \Big(m v\mdot \veps \cosh \left(k\mdot a\right)+i k\mdot S\mdot \veps\frac{ \sinh \left(k\mdot a\right) }{k\mdot a}\Big)\, .
    \end{align}  
\section{Singularities in general spinning hypersurface}
Here we study the physical effects arising from the Kerr action in eq.~\eqref{eq:actionMSKerr} and the corresponding amplitudes. For comparison, we allow the lower-dimensional couplings involving the Riemann tensor to have arbitrary coefficients $\xi_{1,2,4}$. We find that only the Kerr action yields the simple-pole divergence in the bending angle of the scattering with a test particle and maintains an unaltered stable vacuum solution of $Z^\mu_{\mathbf{0}}$, in contrast to an action with generic $\xi_{1,2,4}$. 
 The bending angle is computed at the first post-Minkowskian order with impact parameter $b$ in the ``aligned-spin'' configuration $v_1\mdot a=v_2\mdot a=0$ (where $v_1,v_2$ denote the velocities of the spinless test particle and the spinning black hole, respectively).
We have explored a large number of terms in the general action \eqref{eq:actionMS} and for generic $\xi_{1,2,4}$, the spin-resummed bending angle at this order is 
\begin{align}
	\chi^{(1,1)}_{1\rm PM}(\xi_1,\xi_2,\xi_4)&=\sum_{i=1}^5c_{i}\chi^{(1,1)}_{1\rm PM}(B_i)
\end{align}
where $\chi^{(1,1)}_{1\rm PM}(B_i)$ denotes the contribution from the basis element $B_i$. We find that the singularities always lie at $|a|=|b|$. For the Kerr action in eq.~\eqref{eq:actionMSKerr}, only $B_1$ and $B_4$ contribute and the result agrees with~\cite{Newman:1965tw,Guevara:2019fsj,Brandhuber:2023hhl,Arkani-Hamed:2019ymq,Kim:2024grz,Chen:2024bpf}. For generic $\xi_{1,2,4}$, the resulting bending angle has logarithmic divergences. Further analysis of the bending angle involving higher modes is presented in the \emph{Supplemental Material}~\cite{SuppMat}.

The spin-resummed bending angle at the second post-Minkowskian order is presented in the \emph{Supplemental Material}~\cite{SuppMat}.  Similarly, we find that the singularities always lie at $|a|=|b|$. The second post-Minkowskian bending angle for Kerr agrees with \cite{Damgaard:2022jem,Chen:2024bpf}, also free of logarithmic divergences. For generic $\xi_{1,2,4}$, the existence of logarithmic divergences persists. 

Another critical direction to consider is the metric, given the hypersurface's interaction with gravity. 
We start with the Einstein equation expanded up to the first post-Minkowskian order,
 \begin{align}\label{eq:EinsteinEq}
	&R^{\mu\nu}-{R\over 2}g^{\mu\nu}= 8\pi G_N T^{\mu\nu}=8\pi G_N{-2\over \sqrt{-g}}{\delta S\over \delta g_{\mu\nu}(x)} \, ,
\end{align} 
where 
\begin{align*}
\begin{split}
g_{\mu\nu}(x)&=\eta_{\mu\nu}+\kappa^2 \mathfrak h_{\mathbf{1}\,\mu \nu }(x)+\mathcal{O}(\kappa^4)\\
T^{\mu \nu }(x)&=T_{\mathbf 0}^{\mu \nu }(x)+\mathcal{O}(\kappa)
\end{split}\, .
\end{align*}
At this order, $\mathfrak h_{\mathbf{1}\,\mu \nu }$ contains only the leading-order stationary contribution. Fourier transforming eq.~\eqref{eq:EinsteinEq} to momentum space (see \cite{Mougiakakos:2020laz,Damgaard:2024fqj,Mougiakakos:2024nku} for spinless examples), we find it is directly related to $A_3^{(1,1)}$ \footnote{We thank Yuexiang Zhang for discussions on this point and for sharing his calculations for cross-checking.},
\begin{align}
\widetilde{\mathfrak h}_{\mathbf{1}}^{\mu \nu }(q)&=\frac{ \widetilde T_{\mathbf{0}}^{\mu \nu }(q)-{1\over 2}\widetilde T^{\rho}_{\mathbf{0}\rho}(q) \eta^{\mu \nu }}{2 q^2}\nn\\
&={2\pi \delta(m v\mdot q)\over 2q^2}\sum_{\rm spin~ sum} {A_3^{(1,1)}(v,a, q)\over -i\kappa}\veps^{*\mu}\veps^{*\nu}\,.
\end{align}
Here we work in the de Donder gauge, and $\veps^{*}$ denotes the conjugate of $\veps$. 
The first post-Minkowskian stationary metric in position space is then given by 
\begin{align}
       \mathfrak h^{\mu\nu}_{\mathbf{1}}(x)=\int {d^{4}q\over (2\pi)^4} e^{-i  q\mdot x}\widetilde{\mathfrak h}_{\mathbf{1}}^{\mu \nu }(q)\, .
\end{align}       
Explicit expressions are listed in tab.~\ref{tab:1}. The metric always satisfies the Einstein equation above for any $\xi_{1,2,4}$. 

As shown in tab.~\ref{tab:1}, entire functions $B_1, B_4$ that are present in the Kerr amplitude give rise to the ring singularity in the metric, while the rest ($B_2, B_3, B_5$ in eq.~\eqref{eq:A311}) contribute only to the regular part of the metric at first post-Minkowskian order.   
Interestingly, the generated Kerr metric $\mathfrak{h}_{\mathbf{1}}^{\mu\nu}$ (only $B_1+B_4$ in tab.~\ref{tab:1}) vanishes as $r\rightarrow 0$ if the ring singularity is appropriately regularised, suggesting the flat-space solution of $Z_{\mathbf{0}}^\mu$ remains stable and unaffected by $\mathfrak{h}_{\mathbf{1}}^{\mu\nu}$ at this order. 
For a more general action, $\mathfrak{h}_{\mathbf{1}}^{\mu\nu}$ receives contributions from $B_2, B_3, B_5$, which do not vanish as $r\rightarrow 0$ and can modify the flat-space solution. This observation may corroborate the stability of the Kerr black hole from a novel viewpoint. 
\section{Conclusion and outlook}
This letter takes a first step in exploring a new hypersurface model, in which the action preserves diffeomorphism invariance across the world volume and the target spacetime, yielding solutions that rely exclusively on variables such as velocity, classical spin, and mass.   We note that the spin tensor arises naturally from the rotational angular momentum of the Kerr disc \cite{Israel:1970kp} around its center of mass (In contrast, the worldsheet action in \cite{Guevara:2020xjx} introduces the spin degree of freedom extrinsically via the Newman-Janis shift). As we have observed, this framework effectively encapsulates the external states necessary for an in-depth characterization of the Kerr black hole, offering interesting avenues for further study. The pivotal result of our analysis is the successful generation of the classical three-point Kerr amplitude to arbitrary spin order directly from the model's equation of motion. We hope that this constructive result will enhance the theoretical precision of the computational analysis of observations of binary events.

For instance, by expanding the action's relative coordinate, we can distill the action into a worldline formulation comprising an infinite number of terms beyond minimal coupling. While our hypersurface action involves only metric and Riemann tensor couplings, derivatives of the Riemann tensor and multiple Riemann tensor couplings naturally emerge from this expansion, opening intriguing new avenues for worldline computations. 

Within the framework established for the three-point amplitude, it is crucial to investigate how to use this action to derive high-multiplicity classical amplitudes by solving the corresponding extended equations of motion. The major task is to solve both the hypersurface coordinate equation of motion and the Einstein equation eq.~\eqref{eq:EinsteinEq}. 
The above equation of motion can be solved perturbatively in both spin parameter $|a|$ and gravity coupling $\kappa$ order by order. One can thus extract the high-multiplicity amplitude perturbatively from the outgoing gravitational wave solution in the presence of multiple incoming plane waves or partial waves (see {\it i.e.}~\cite{Scheopner:2023rzp} for related discussions on Compton amplitude). The advantage of the hypersurface equation of motion is that it obviates the need for a multipole expansion or the introduction of multipole moments, while still yielding an amplitude that depends only on the spin and mass of the black hole. We leave a detailed discussion to an upcoming paper (see also the illustrative discussions in the \emph{Supplemental Material}~\cite{SuppMat}). To verify the obtained amplitude, we examine the factorization behavior and use the model's contact terms to gain new insights, for example, in the context of existing Compton amplitude bootstrap techniques \cite{Bjerrum-Bohr:2023jau, Bjerrum-Bohr:2023iey, Chen:2024mlx}. 

Given that our framework encompasses intricate characteristics of extended bodies, it also facilitates the extraction of the multipole expansion from the energy-momentum tensor (see {\it i.e.}~\cite{Scheopner:2023rzp} using disc source). It opens a new avenue for analytically addressing the Mathisson-Papapetrou-Dixon equations within the post-Minkowskian expansion framework and for conducting numerical analyses in scenarios characterized by strong gravitational coupling. Such investigations go beyond the scope of this presentation. Still, they are vital as they will likely serve as future cross-validation of the theoretical framework we envision for higher-multiplicity amplitudes. 

Besides identifying a correspondence with the amplitude associated with a Kerr black hole, we emphasize that the proposed hypersurface model for spinning black holes extends to a broad class of interactions in arbitrary dimensions, potentially leading to a new understanding of the complex behavior of extended bodies that could be relevant for future astrophysical phenomenology and precision observations. 

\begin{acknowledgements}
{\bf Acknowledgements}
We thank Andreas Brandhuber, Paolo Di Vecchia, Yutin Huang, Jung-Wook Kim, and Marcos Skowronek for discussions. The work of N.E.J.B.-B. and G.C. was supported in part by DFF grant 1026-00077B, and by The Center of Gravity, which is a Center of Excellence funded by the Danish National Research Foundation under grant No. 184. G.C. has also received funding from the European Union Horizon 2020 research and innovation program under the Marie Sklodowska-Curie grant agreement No. 847523 INTERACTIONS, as well as support in part by VILLUM Foundation (grant no. VIL37766). Guangzhou Elite Project supported C.S. through grant number JY202211. T.W. was supported by the National Research Foundation of Korea (NRF) grant NRF RS-2024-00351197.
\end{acknowledgements}

\bibliographystyle{apsrev4-1}

\bibliography{KinematicAlgebra}
\newpage
\appendix 
\newpage
\onecolumngrid
\begin{center}
    \textbf{\large Supplemental material for  \\``Kerr Black Hole Dynamics from an Extended Polyakov Action''}
\end{center}
In this supplemental material, we outline the computation of the gravitational Compton amplitude at leading order using the proposed action, thereby establishing a concrete path for computing higher-point amplitudes from the worldvolume model. We also present a discussion, which complements the main text, addressing the possibilities of alternative worldvolume metrics. % and the stability of the three-point amplitude under superpositions of external states. 
Finally, we provide further details on the computations of the bending angle derived from the $Y_{1,1}$-mode at the second post-Minkowskian order and list the first post-Minkowskian bending angle results for several higher modes.  \\ ~ \\
\twocolumngrid

\section{Gravitational Compton amplitude}
We will here outline the computation of the gravitational Compton amplitude of a Kerr black hole from the model up to $\mathcal O(|a|)$. The generalization of the method to higher powers of spin is straightforward, but involves additional bookkeeping and techincal steps, and it is thus beyond the scope of this letter. We plan to study it further in our upcoming publication.\\[5pt]
The gravitational Compton amplitude of a Kerr black hole in our model can be extracted from the solutions to the Einstein equation with the stress-energy tensor,
\begin{align}\label{eq:Tmunu}
	T^{\mu\nu}(x) =  & {m\over 4\pi a_w^2\sqrt{-g}}\int d^3\sigma \sqrt{\gamma}\delta^{4}(x-Z) (\partial_a Z^\mu \gamma^{ab}\partial_b Z^\nu) \nonumber\\
	& + \dots\,,
\end{align} 
where we only write down the contribution from the Polyakov term in the action $S_{\rm Kerr}$, as this is sufficient for the computation through $\mathcal {O} (|a|)$.
We next define the metric around flat space $g_{\mu\nu}\equiv\eta_{\mu\nu}+\kappa \mathrm{h}_{\mu\nu}$, and expand the perturbation field $\mathrm{h}_{\mu\nu}$ in increasing orders of $\kappa$ as follows 
\begin{align}
	\mathrm{h}_{\mu\nu}&=h_{\bz\, \mu\nu}+\kappa\, h_{\bo\, \mu\nu}+\kappa^2\, h_{\mathbf{2}\,\mu\nu}+\cdots\,.
\end{align} 
The stress-energy tensor is expanded similarly 
\begin{align}
T^{\mu\nu}=T^{\mu\nu}_{\mathbf{0}}+\kappa\, T^{\mu\nu}_{\mathbf{1}}+\cdots \, .
\end{align}
We take $h_{\bz\, \mu\nu}$ to denote an incoming graviton taken to be a plane wave represented by
\begin{align}
h_{\bz\,\mu\nu}=\lambda\, \veps_{1\mu}\veps_{1\nu} e^{-ik_1\mdot x}, 
\end{align}
where $\lambda$ has been introduced as a convenient scale factor accompanying the incoming graviton wave polarization vector, as our tactic is to compute the Compton amplitude by the $\mathcal{O}(\lambda)$ response to the interaction with the black hole. The relevant contributions (to the order we work) can be outlined as follows, $h_{\mathbf{1}\,\mu\nu} = \mathfrak h_{\mathbf{1}\,\mu\nu}+\mathcal{O}(\lambda^2)$, and $h_{\mathbf{2}\,\mu\nu} = \mathfrak h_{\mathbf{2}\,\mu\nu} + \lambda \delta h_{\mathbf{2}\,\mu\nu}+\mathcal{O}(\lambda^2)$, where $\mathfrak{h}_{\mathbf{i}}$ denotes a stationary contribution and $\delta h_{\mathbf{2}}$ the effect arising from an outgoing wave contribution.

We next expand the Einstein equation at $\mathcal{O}(\kappa^3\lambda)$ in the de Donder gauge $\partial_\mu h_{\mathbf{i}}^{\mu\nu}-{1\over 2} \partial^{\nu}h^{\mu}_{\mathbf{i}\,\mu}=0$, and arrive at
\begin{align}\label{eq:Einsteinkappa3}
&-\partial^2 \delta h^{\mu\nu}_{\mathbf 2}+\eta^{\mu\nu}\partial_{\rho}\partial_{\sigma}\delta h^{\rho\sigma}_{\mathbf{2}} =  
    \frac{T_{\mathbf{1}}^{\mu\nu}}{2} 
    - M_{\mathbf{1}}^{\mu\nu}.
\end{align}
%
%Here $T_{\mathbf{1}}^{ \mu\nu}$ denotes the stress-energy tensor at $\mathcal{O}(\kappa\lambda)$ and 
Here $M_{\mathbf{1}}^{\mu\nu}$ denotes the graviton scattering from the stationary metric (written in conventional index notation) at $\mathcal{O}(\kappa\lambda)$, given by
\begin{align}\label{M1 in coordinate}
  &M_{\mathbf{1}}^{\mu\nu}= \frac{1}{2}h_{\bz}^{\rho\sigma}\partial_{\rho}\partial_{\sigma}\mathfrak h_{\bo}^{\mu\nu}+\frac{1}{2}(\partial^{(\mu}h_{\bz\,\rho\sigma})\partial^{\nu)}\mathfrak h_{\bo}^{\rho\sigma}\nn\\
    &-(\partial_{\rho}h_{\bz}^{\sigma(\mu})\partial_{\sigma}\mathfrak h_{\bo}^{\nu)\rho}+h_{\bz}^{\rho(\mu}\partial^2\mathfrak h_{\bo\rho}^{\nu)}+h_{\bz}^{\rho(\mu}\partial^{\nu)}\partial_{\sigma}\mathfrak h_{\bo\rho}^{\sigma}\nn\\
    &-\frac{1}{2}h_{\bz}^{\mu\nu}\partial_{\rho}\partial_{\sigma}\mathfrak h_{\bo}^{\rho\sigma}+\frac{1}{2}h_{\bz}^{\rho\sigma}\partial^{\mu}\partial^{\nu}\mathfrak h_{\bo\rho\sigma}+(\partial^{\rho}h_{\bz}^{\sigma(\mu})\partial_{\rho}\mathfrak h_{\bo\sigma}^{\nu)}\nn\\
    &-h_{\bz}^{\rho(\mu}\partial_{\rho}\partial_{\sigma}\mathfrak h_{\bo}^{\nu)\sigma}-h_{\bz}^{\rho\sigma}\partial_{\rho}\partial^{(\mu}\mathfrak h_{\bo\sigma}^{\nu)}+\frac{1}{2}(\partial^{\mu}\partial^{\nu}h_{\bz}^{\rho\sigma})\mathfrak h_{\bo\rho\sigma}\nn\\
    &-h_{\bz}^{\rho(\mu}\partial^{\nu)}\partial_{\sigma}\mathfrak h_{\bo\rho}^{\sigma}+\frac{1}{2}(\partial_{\rho}\partial_{\sigma}h_{\bz}^{\mu\nu})\mathfrak h_{\bo}^{\rho\sigma}+ (\cdots)
\end{align}
We have omitted the explicit expressions of terms containing $\eta^{\mu\nu}, \partial_\rho h_{\bz}^{\rho\sigma}$ factors above, as they do not lead to any relevant contributions. From the right-hand side of eq.~\eqref{eq:Einsteinkappa3} we will derive the gravitational Compton amplitude, with an incoming graviton of momentum $k_1$ and outgoing momentum $k_2$, and spin $a$ and heavy-mass velocity $v$. We start by Fourier transforming (\ref{eq:Einsteinkappa3}) to the momentum space of the outgoing wave.  Contracting indices with the required polarization vectors and adjusting the global factor of $\kappa$ (see {\it e.g.} \cite{Scheopner:2023rzp}), it leads to the following expression for the Compton amplitude in our notation
 \begin{align}
& A_4(v,a,k_1,k_2) ={\kappa^2\over \lambda} \Big(
    -\frac{\widetilde T_{\mathbf{1}}^{\mu\nu}}{2} \veps_{2\mu}\veps_{2\nu}
    +  \widetilde M_{\mathbf{1}}^{\mu\nu}\veps_{2\mu}\veps_{2\nu}
\Big) \, .
\end{align}
Here, the tildes provide a shorthand notation for Fourier-transformed quantities,
\begin{align}
\widetilde{(\bullet)}(k_2)&=\int d^4 x e^{ik_2\mdot x}{(\bullet)}(x)\, .
\end{align}
$\widetilde M_{\mathbf{1}}^{\mu\nu}(k_2)$ is now readily obtained by simple manipulations, for example
\begin{align}
&  \int d^4 x e^{ik_2\mdot x}\Big({1\over 2}h_{\bz}^{\rho\sigma}\partial_{\rho}\partial_{\sigma}\mathfrak h_{\bo}^{\mu\nu}\Big)\nn\\
&={\lambda\over 2}\int d^4 x e^{i(k_2-k_1)\mdot x}( \veps_1^{\rho}\veps_1^{\sigma})\partial_{\rho}\partial_{\sigma}\mathfrak h_{\bo}^{\mu\nu}(x) \nn\\
& ={\lambda\over 2}(-i(k_{2}\mdot\veps_1-k_{1}\mdot\veps_1))^2\ \widetilde{\mathfrak h}_{\bo}^{\mu\nu}(k_2-k_1) \, ,
\end{align}
where ${\widetilde {\mathfrak h}}^{\mu\nu}_{\bo}(k_2-k_1)$ is the $\mathcal{O}(\kappa^2)$ stationary metric in momentum space.\\[5pt]
To evaluate the Fourier-transformed stress-energy tensor 
\begin{align}\label{eq:Tmomentum}
\widetilde T_{\mathbf{1}}^{\mu\nu}(k_2)&= {\lambda \over 4\pi a_w^2}\int d^3\sigma \sqrt{\gamma}e^{i k_2\mdot Z_{\mathbf{0}}}\\
& \times\Big[i k_2\mdot Z_{\mathbf{1}}\partial_a Z_{\mathbf{0}}^{\mu} \gamma^{ab}\partial_b Z_{\mathbf{0}}^{\nu}+2\gamma^{ab} \partial_a Z_{\mathbf{0}}^{(\mu} \partial_b Z_{\mathbf{1}}^{\nu)}\Big]\, ,\nn
\end{align}
 we first solve for the next-to-leading order correction to the spacetime coordinate $Z^\rho(\sigma)$. The equation of motion for $Z^\rho$ follows from the variation of the terms in the action $S_{\rm Kerr}$ that are required to the order we work, 
 \begin{align}\label{eq:EOMGenNew}
	\partial_a \Big(\sqrt{\gamma}\gamma^{ab}\partial_b Z^\rho \Big)+\sqrt{\gamma}\partial_a Z^\mu \gamma^{ab}\partial_b Z^\nu { \Gamma_{\mu\nu}^\rho(Z)}=0\,.
\end{align}
Expanding $Z^\mu(\sigma)$ we write
\begin{align}
	Z^{\mu}(\sigma)&=Z^{\mu}_{\mathbf{0}}(\sigma)+\kappa \lambda Z^{\mu}_{\mathbf{1}}(\sigma)+\cdots  \, ,
\end{align}
where $Z_{\mathbf{0}}^\mu\equiv X^\mu_{\mathbf{0}}(\tau)+Y^\mu_{1,1}(\tau,\theta,\phi)$ denotes the solution to the flat-space equation of motion given in the main text.\\[5pt]
The background metric evaluated on the worldvolume admits the following post-Minkowskian expansion,
	\begin{align}
	\mathcal{G}_{\rho\nu}(Z)&=\eta_{\rho\nu}+\kappa\lambda \veps_{1\rho}\veps_{1\nu} e^{-ik_1\mdot Z}+\cdots \, .
	\end{align}
At $\mathcal O(\kappa\lambda)$, using $Z^\rho_{\mathbf{1}}=e^{-ik_1\mdot X_{\mathbf{0}}}\bar Z^\rho_{\mathbf{1}}$, we arrive at the equation of motion
\begin{align}\label{eq:EOMGenZbar}
	&\!\!\partial_a \!\Big(\!\sqrt{\gamma}\gamma^{ab}\partial_b \bar Z^\rho_{\mathbf{1}} \Big)\!-\!a_w^2 \sin(\theta)\big((k_1\mdot v)^2 \bar Z^\rho_{\mathbf{1}}\!+\!2i  k_1\mdot v \partial_\tau \bar Z_1^\rho\big)\!\!\!\! \!\!\!\! \\
    &\!\! =
    i e^{-ik_1\mdot Y_{\mathbf{0}}}\!\sqrt{\gamma}\gamma^{ab}\partial_b (\veps_1\mdot Z_{\mathbf{0}})\Big(  \partial_a(k_1\mdot Z_{\mathbf{0}})\veps_1^\rho\!-\!{1\over 2}\partial_a (\veps_1\mdot Z_{\mathbf{0}})  k_1^{\rho} \Big)\, .\nn
\end{align}
We can further decompose $\bar Z_{\mathbf{1}}^\rho$ into the complete orthonormal basis of spherical harmonic functions $\mathcal{Y}_{l,m}(\theta,\phi)$,
\begin{align}
\bar Z^{\rho}_{\mathbf{1}}(\sigma)= \sum_{l\geq |m|=0}^{\infty}\Psi^{\rho}_{\mathbf{1},l,m }(\tau)\mathcal{Y}_{l,m}(\theta,\phi)\, .
\end{align}
It provides the following second-order differential equation for a given mode $\mathcal{Y}_{l,m}$ and  its coefficients $\Psi_{\mathbf{1},l,m}(\tau)$,
\begin{align}
 &2 a_w^2 \left(\partial^2_\tau\Psi _{\mathbf{1},l,m}^{\rho }(\tau )-i k_1\cdot v_1 \partial_\tau\Psi _{\mathbf{1},l,m}^{\rho }(\tau )\right) \nn\\
 &+2\left(l(l+1)- a_w^2 \left(k_1\mdot v_1\right){}^2\right) \Psi _{\mathbf{1},l,m}^{\rho }(\tau )&\nn\\
 &=\int_0^\pi d\theta \int_0^{2\pi}d\phi  \mathcal{Y}^{*}_{l,m}(\theta,\phi)\big[\text{r.h.s. of \eqref{eq:EOMGenZbar}}\big].
 \end{align}
Boundary conditions are chosen consistent with expectations of a physical scattering process of a hypersurface in an asymptotically flat, curved background, i.e. such that the solution to the equation of motion approaches the flat-space one once the incoming graviton has passed. It allows us to perturbatively solve the coefficients $\Psi^\rho_{\mathbf{1},l,m}(\tau)$ analytically for different powers of $a_w$. They consist of sums of different frequencies, which can be easily identified as integer powers of $\sqrt{2}/a_w$. Hence, $Z_{\mathbf{1}}^\rho$ has to take the form
\begin{align}
Z^\rho_{\mathbf{1}}= e^{-ik_1\mdot X_\mathbf{0}}\sum_{n,l,m}C^{\rho}_{n,l,m} e^{i {\sqrt{2} n\over a_w}\tau} \mathcal{Y}_{l,m}(\theta,\phi)\, .
\end{align}
and working to order $\mathcal{O}(a_w)$, the solutions of the relevant $\Psi_{\mathbf{1},l,m}(\tau )$'s read
\begin{align}
&\Psi_{\mathbf{1},0,0}^{\rho}(\tau )= \frac{i\sqrt{\pi } v_1\mdot \veps _1 \left(k_1^{\rho } v_1\mdot \veps _1-2 k_1\mdot v_1 \veps _1^{\rho }\right)}{\left(k_1\mdot v_1\right){}^2}-\frac{2  \sqrt{2 \pi } c_{1,1}^2 a_w }{3 \left(k_1\mdot v_1\right){}^2}\nn\\
&\times \left(k_1^{\rho } v_1\mdot \veps _1-k_1\mdot v_1 \veps _1^{\rho }\right) \left(k_1\mdot \beta _y \beta _x\mdot \veps _1-k_1\mdot \beta _x \beta _y\mdot \veps _1\right)+\mathcal{O}(a_w^2)\nn\\
&\Psi_{\mathbf{1},1,\pm 1}^{\rho}(\tau )= i \sqrt{\frac{ \pi }{6}} c_{1,1} a_w e^{\pm \frac{i \sqrt{2} \tau }{a_w}}\Big( \veps _1^{\rho } \left(\beta _y\mdot \veps _1\pm i \beta _x\mdot \veps _1\right)\!\!\!\!\!\!\!\! \\ \!\!\!\!
	&\!+\!\frac{ v_1\mdot \veps _1 \left(\veps _1^{\rho } \left(k_1\mdot \beta _y\!\pm\!i k_1\mdot \beta _x\right)\!-\!k_1^{\rho }( \beta _y\mdot \veps _1\!\pm\!i  \beta _x\mdot \veps _1)\right)}{k_1\mdot v_1}\Big)\!+\!\mathcal{O}(a_w^2)\nn \, .
\end{align}
Therefore,  
\begin{align}\label{eq:solZ1}
Z^\rho_{\mathbf{1}}&= e^{-ik_1\mdot X_\mathbf{0}}\Big(\Psi^{\rho}_{\mathbf{1},0,0}(\tau ) \mathcal{Y}_{0,0}(\theta,\phi)+\Psi^{\rho}_{\mathbf{1},1,1}(\tau ) \mathcal{Y}_{1,1}(\theta,\phi)\nn\\
&+\Psi^{\rho}_{\mathbf{1},1,-1}(\tau ) \mathcal{Y}_{1,-1}(\theta,\phi)\Big)\, +\mathcal{O}(a_w^2).
\end{align}
Substituting the solutions of $Z_{\mathbf{0}}^\rho$ and $Z_{\mathbf{1}}^\rho$ and carrying out the remain integrals, we arrive at
\begin{align}
\widetilde{T}^{\mu\nu}_{\mathbf{1}}\veps_{2\mu}\veps_{2\nu}&=\lambda\, \delta(v\mdot k_2\!-\!v\mdot k_1) \bigg[\Big[{{v_1\mdot \veps _1 v_1\mdot \veps _2\over 2(k_1\mdot v_1)^2}\Big(2 v_1\mdot F_1\mdot F_2\mdot v_1}\nn\\
& \!\!\!\!\!\! \! \! \! \! \! \! \! \! \! \!\! \! \! \! \! \! \! \! \! \!  {-2 \veps _1\mdot \veps _2  \left(k_1\mdot v_1\right){}^2+k_1\mdot k_2 v_1\mdot \veps _1 v_1\mdot \veps _2\Big)}\\
&\!\!\!\!\!\! \! \! \! \! \! \! \! \! \! \!\! \! \! \! \! \! \! \! \! \! {-{i\over 2m(k_1\mdot v_1)^2}\Big(v_1\mdot \veps_2 k_1\mdot S\mdot \veps_1(-2 v_1\mdot F_1\mdot F_2\mdot v_1}\nn\\
&\!\!\!\!\!\! \! \! \! \! \! \! \! \! \! \!\! \! \! \! \! \! \! \! \! \! {+k_1\mdot v_1 k_2\mdot \veps _1 v_1\mdot \veps _2-k_1\mdot k_2 v_1\mdot \veps _1 v_1\mdot \veps _2)+(1\leftrightarrow 2)\Big)\Big]}\nn\\
&\!\!\!\!\!\!\! \! \! \! \! \! \! \! \! \! \!\! \! \! \! \! \! \! \! \! \! +\Big[\frac{i}{2m}  \Big(-\frac{v_1\mdot \veps _1 v_1\mdot \veps _2 \text{tr}\left(F_1\mdot F_2\mdot S\right)}{k_1\mdot v_1}-v_1\mdot \veps _2 \veps _1\mdot F_2\mdot S\mdot \veps _1\nn\\
&\!\!\!\!\!\!\! \! \! \! \! \! \! \! \! \! \!\! \! \! \! \! \! \! \! \! \! +v_1\mdot \veps _1 \veps _2\mdot F_1\mdot S\mdot \veps _2\!-\!\veps _1\mdot \veps _2 k_1\mdot v_1 \veps _1\mdot S\mdot \veps _2\Big)\!  \Big]\bigg]\nn\\ &\!\!\!\!\!\!\! \! \! \! \! \! \! \! \! \! \!\! \! \! \! \! \! \! \! \! \! +\mathcal{O}(a_w^2),\nn
\end{align}
where $F_i^{\mu\nu}=k^\mu_i\veps^\nu_i-\veps^\mu_i k^\nu_i$. We note that the terms in the first square bracket are due to the mode $\mathcal{Y}_{0,0}$, while the terms in the second arrive from the mode $\mathcal{Y}_{1,\pm 1}$.\\[5pt]
Using the expression for $\widetilde M_{\bo}$, we finally obtain the gravitational Compton amplitude from our model
\begin{align}
&A_4(v,a,k_1,k_2)=\kappa^2\delta(v\mdot k_2-v\mdot k_1)\\
&\times\Big[ \Big({v\mdot F_1\mdot F_2\mdot v\over v\mdot k_1}\Big)^2-\frac{i v\mdot F_1\mdot F_2\mdot v }{2 k_1\mdot k_2 \left(v\mdot k_1\right){}^2}\nn\\
&\times\Big(v\mdot k_1 \text{tr}\left(S\mdot F_2\mdot F_1\right)+v\mdot F_1\mdot F_2\mdot S\mdot k_2+k_1\mdot S\mdot F_1\mdot F_2\mdot v\Big) \Big]. \nn
\end{align}
This result for the Compton amplitude agrees with the literature, see e.g. \cite{Bjerrum-Bohr:2023jau}. 

\section{On the possibilities of alternative worldvolume metrics}
Reparameterisation symmetry on the worldvolume does allow us to start with a more general diagonal form of the worldvolume metric $ds^2 = f_1 d\tau^2-f_2 d\theta^2-f_3 d\phi^2$ where we can define $f_i\equiv f_i(\tau,\theta,\phi)$ to be generic functions of the coordinates $\tau$, $\theta$, and $\phi$. However, we note, for most choices of $f_i$'s, the solution to the equation of motion $\partial_a\!\big(\sqrt{\gamma}\,\gamma^{ab}\partial_bZ^{\mu}\big)= 0 $ results in a very complicated differential equation, with solutions typically not expressible in terms of simple elementary functions. To solve for the corresponding spin-expanded three-point amplitude, one further has to evaluate the integral over the spatial worldvolume coordinates,
\begin{align}
\int d\theta d\phi \sqrt{\gamma} (\partial_{b_1} Y  \cdots\partial_{b_i} Y) (Y)^j\,,
\end{align}
and after exploring a wide variety of potential alternative candidate metrics, we have reached a preliminary conclusion that sensible results are only found for metrics of the type.
\begin{align}
\mathbb{S}_1\times \mathbb{S}_1\times \mathbb{R}:~&ds^2 = dt^2-a_w^2\, d\theta^2 - a_w^2 \, d\phi^2\,,\nn\\
\mathbb{S}_2\times \mathbb{R}:~&ds^2 = dt^2-a_w^2\, d\theta^2 - a_w^2 \sin^2{\theta}\, d\phi^2\,,
\end{align}
(e.g., one particular problem is that non-elementary $Y^\mu$ solutions leads to coefficients with powers of $(k\mdot a)$ in the three-point amplitude being non-rational numbers. Such solutions we discarded as possible candidate metrics). We have further explored the metric choice $\mathbb{S}_1 \times \mathbb{S}_1 \times \mathbb{R}$, which leads to Bessel and elliptic functions after resumming the spin expansion. For the Kerr black hole amplitude, the only sensible choice we have identified at the time of writing is the $\mathbb{S}_2 \times \mathbb{R}$, which we explore in the main text. 

\section{Spin-Resummed Bending Angle}
Here we present the spin-resummed bending angle at the second post-Minkowskian order computed from the worldvolume action with generic parameters evaluated for the $Y_{1,1}$-mode. The bending angle can be expressed as a sum over products of two basis elements 
\begin{align}
	\chi^{(1,1)}_{\rm 2PM}&=\sum_{1\leq i\leq j\leq 5}c_{i}c_j\chi^{(1,1)}_{\rm 2PM}(B_i\otimes B_j),
\end{align}
where the basis elements denoted as $B_i$ are entire functions given in the main text and $B_i\otimes B_j$ label combinations. We list all the resummed bending angles for each entire function's basis in the tab.~\ref{tab:Eiko2PM}.\\[5pt]
To provide the Kerr solution, the spin-resummed second post-Minkowskian bending angle has to be given by
$\chi^{(1,1)}_{\rm 2PM}(B_1\otimes B_1)
+\chi^{(1,1)}_{\rm 2PM}(B_4\otimes B_4)+ \chi^{(1,1)}_{\rm 2PM}(B_1\otimes B_4)$,
which is in agreement with \cite{Damgaard:2022jem} and has power-law singularities only as expected \cite{Newman:1965tw,Guevara:2019fsj,Brandhuber:2023hhl,Arkani-Hamed:2019ymq,Kim:2024grz,Chen:2024bpf}. For non Kerr solutions, we note that logarithmic singularities arise at $|a|=|b|$.
\begin{table*}
    \centering
    \begin{tabular}{c|l}
    &$\chi_{\rm 2PM}^{(1,1)}(B_i\otimes B_j)/(\frac{\kappa ^4 m_2 \sqrt{s}}{1024 \pi  \left(y^2-1\right)^2|b|^2} )$ \\ \hline \hline \\ [-9pt]
       $B_1\otimes B_1$  & $\frac{\sqrt{1-\zeta ^2} \left(-2 \left(y^2-1\right)^2+3 \zeta ^4 \left(5 y^4-2 y^2+1\right)+\zeta ^2 \left(35 y^4-46 y^2+11\right)\right)}{8 \zeta ^2 \left(\zeta ^2-1\right)^3 }-\frac{\left(y^2-1\right)^2}{4 \zeta ^2 }$ \\ \hline\\   [-9pt]
$B_1\otimes B_2$  &$ -\frac{\sqrt{1-\zeta ^2} \left(y^2-1\right)^2}{\zeta ^2 \left(\zeta ^2-1\right)}-\frac{\zeta ^4+\left(\zeta ^4+6 \zeta ^2+1\right) y^4-2 \left(3 \zeta ^2+1\right) y^2+1}{\zeta ^2 \left(\zeta ^2-1\right)^2}$ \\ \hline\\   [-9pt]
$B_1\otimes B_3$  &
$\frac{\log (1-\zeta ) \left(-8 \zeta  \left(y^2-1\right)^2-6 y^2+3\right)}{8 \zeta ^3}+\frac{\log (\zeta +1) \left(-8 \zeta  \left(y^2-1\right)^2+6 y^2-3\right)}{8 \zeta ^3}+\frac{2 \left(y^2-1\right)^2 \log \left(\sqrt{1-\zeta ^2}+1\right)}{\zeta ^2}$\\
&$-\frac{\left(y^2-1\right)^2 \log (4)}{\zeta ^2}+\frac{-6 y^2+4 \zeta ^4 \left(y^4-3 y^2+1\right)+\zeta ^2 \left(-12 y^4+26 y^2-9\right)+3}{4 \zeta ^2 \left(\zeta ^2-1\right)^2}$ \\ \hline\\   [-9pt]
$B_2\otimes B_2$  & $\frac{\left(1-2 y^2\right)^2}{\zeta ^2-1}-\frac{y^2 \log (1-\zeta )}{\zeta ^2}-\frac{y^2 \log (\zeta +1)}{\zeta ^2}+\frac{\left(y^2-1\right)^2 \log (2)}{\zeta ^2}-\frac{\left(y^2-1\right)^2 \log \left(\sqrt{1-\zeta ^2}+1\right)}{\zeta ^2}$ \\ \hline\\   [-9pt]
$B_2\otimes B_3$  &
$\frac{8 \zeta ^2+4 \left(5 \zeta ^2-4\right) y^4+\left(18-22 \zeta ^2\right) y^2-7}{4 \zeta ^2 \left(\zeta ^2-1\right)}+\frac{\log (1-\zeta ) \left(\zeta  (3-4 \zeta )-4 \left(\zeta ^2-8\right) y^4-6 (\zeta +4) y^2+8\right)}{8 \zeta ^4}+\frac{\log (\zeta +1) \left(-\zeta  (4 \zeta +3)-4 \left(\zeta ^2-8\right) y^4+6 (\zeta -4) y^2+8\right)}{8 \zeta ^4}$\\ \hline\\   [-9pt]
$B_3\otimes B_3$  &
$-\frac{\log (1-\zeta ) \left(\zeta ^4-\zeta ^3-4 \zeta ^2+\left(\zeta ^4-16 \zeta ^2+11\right) y^4+2 \left(\zeta ^3+6 \zeta ^2-5\right) y^2+3\right)}{8 \zeta ^6}-\frac{\log (\zeta +1) \left(\zeta ^4+\zeta ^3-4 \zeta ^2+\left(\zeta ^4-16 \zeta ^2+11\right) y^4-2 \left(\zeta ^3-6 \zeta ^2+5\right) y^2+3\right)}{8 \zeta ^6}$\\
&$+\frac{9 \zeta ^2+\left(21 \zeta ^2-22\right) y^4+\left(20-22 \zeta ^2\right) y^2-6}{16 \zeta ^4}$ \\ \hline\\   [-9pt]
$B_4\otimes B_4$&$\frac{\sqrt{1-\zeta ^2} \left(y^2-1\right) \left(3 \zeta ^4-5 \zeta ^2+\left(45 \zeta ^4+5 \zeta ^2-2\right) y^2+2\right)}{8 \zeta ^2 \left(\zeta ^2-1\right)^3}-\frac{\left(y^2-1\right)^2}{4 \zeta ^2}$ \\ \hline\\   [-9pt]
$B_4\otimes B_5$&$\frac{2 \left(y^2-1\right)^2 \log \left(\sqrt{1-\zeta ^2}+1\right)}{\zeta ^2}-\frac{\left(y^2-1\right)^2 \log (1-\zeta )}{\zeta ^2}-\frac{\left(y^2-1\right)^2 \log (\zeta +1)}{\zeta ^2}$\\
&+$\frac{\left(y^2-1\right) \left(\left(1-\zeta ^2\right)^2+\left(3 \zeta ^4-2 \left(2 \sqrt{1-\zeta ^2}+1\right) \zeta ^2-1\right) y^2\right)}{\left(1-\zeta ^2\right)^{5/2} \left(\sqrt{1-\zeta ^2}+1\right)}-\frac{2 \left(-\zeta ^2+\sqrt{1-\zeta ^2}+1\right) \left(y^2-1\right)^2 \log (2)}{\zeta ^2 \sqrt{1-\zeta ^2} \left(\sqrt{1-\zeta ^2}+1\right)}$ \\ \hline\\   [-9pt]
$B_5\otimes B_5$  &
$\frac{\left(y^2-1\right)^2 \log \left(\sqrt{1-\zeta ^2}+1\right)}{\zeta ^2}-\frac{\left(y^2-1\right) \log (1-\zeta ) \left(-\zeta ^2+\left(\zeta ^2+7\right) y^2+1\right)}{2 \zeta ^4}-\frac{\left(y^2-1\right) \log (\zeta +1) \left(-\zeta ^2+\left(\zeta ^2+7\right) y^2+1\right)}{2 \zeta ^4}$\\
&$-\frac{\left(y^2-1\right) \left( \left(\zeta ^2+\left( \zeta ^2-7\right) y^2-1\right)+ \left(\zeta ^2-1\right) \left(y^2-1\right) \log (4)\right)}{2 \zeta ^2 \left(\zeta ^2-1\right)}$ \\ \hline\\ 
$B_1\otimes B_4$& $\frac{2 \zeta  y \sqrt{y^2-1} \left(\left(\zeta ^2+5\right) y^2-3\right)}{\left(1-\zeta ^2\right)^{5/2}}$ \\ \hline\\ 
$B_2\otimes B_4$& 
$-\frac{4 \zeta  y \sqrt{y^2-1} \left(2 \sqrt{1-\zeta ^2}-\zeta ^2 \left(\sqrt{1-\zeta ^2}+1\right)+\left(\left(\sqrt{1-\zeta ^2}+2\right) \zeta ^2-3 \sqrt{1-\zeta ^2}-2\right) y^2+1\right)}{\left(1-\zeta ^2\right)^{5/2} \left(\sqrt{1-\zeta ^2}+1\right)}$\\ \hline\\ 
$B_3\otimes B_4$& 
$\frac{y \sqrt{y^2-1} \left(\zeta ^2+4 y^2-3\right)}{2 \zeta  \left(\zeta ^2-1\right)^2}+\frac{y \left(4 y^2-3\right) \sqrt{y^2-1} \log (1-\zeta )}{4 \zeta ^2}+\frac{y \left(3-4 y^2\right) \sqrt{y^2-1} \log (\zeta +1)}{4 \zeta ^2}$\\ \hline\\ 
$B_1\otimes B_5$&
$\frac{y \left(2 y^2-3\right) \sqrt{y^2-1} \log (1-\zeta )}{2 \zeta ^2}+\frac{y \left(3-2 y^2\right) \sqrt{y^2-1} \log (\zeta +1)}{2 \zeta ^2}$\\ 
&$+\frac{y \sqrt{y^2-1} \left(\zeta ^4+\left(4 \zeta ^2-5\right) \sqrt{1-\zeta ^2} \zeta ^2-4 \zeta ^2+3 \sqrt{1-\zeta ^2}-2 \left(-\sqrt{1-\zeta ^2} \zeta ^2+\sqrt{1-\zeta ^2}+\left(2 \sqrt{1-\zeta ^2}-1\right) \zeta ^4+1\right) y^2+3\right)}{\zeta  \left(\zeta ^2-1\right)^2 \left(\zeta ^2-\sqrt{1-\zeta ^2}-1\right)}$\\ \hline\\ 
$B_2\otimes B_5$&
$\frac{y \sqrt{y^2-1} \log (1-\zeta ) \left(2 (\zeta +4) y^2-3 \zeta \right)}{2 \zeta ^3}+\frac{y \sqrt{y^2-1} \log (\zeta +1) \left(3 \zeta -2 (\zeta -4) y^2\right)}{2 \zeta ^3}+\frac{3 y \left(1-2 y^2\right) \sqrt{y^2-1}}{\zeta  \left(\zeta ^2-1\right)}$ \\ \hline\\ 
$B_3\otimes B_5$&
$-\frac{y \sqrt{y^2-1} \left(-15 \zeta ^2+\left(34 \zeta ^2-28\right) y^2+12\right)}{6 \zeta ^3 \left(\zeta ^2-1\right)}+\frac{y \sqrt{y^2-1} \log (1-\zeta ) \left(-9 \zeta ^3+8 \left(\zeta ^2 (\zeta +3)-7\right) y^2+24\right)}{12 \zeta ^5}+\frac{y \sqrt{y^2-1} \log (\zeta +1) \left(9 \zeta ^3-8 \left((\zeta -3) \zeta ^2+7\right) y^2+24\right)}{12 \zeta ^5}$
   \end{tabular}
   \caption{Spin-resummed contributions to second post-Minkowskian bending angles from the entire function basis. Here $\zeta=|a|/|b|$.}
    \label{tab:Eiko2PM}
\end{table*}

\section{Higher mode examples}
 One can also extend our analysis to higher modes with $l>1,j>1$, where $Y_{l,j}$ is a multi-covering map from $\mathbb{S}^2$ to the disc in the target space. Such three-point amplitudes are still well-defined for a wide range of parameters choices. In particular, for the $(l,l)$- and $(l,l-1)$-modes, the amplitudes can be expressed in terms of the generalized hypergeometric function $_pF_q$ where $p>q+1$ after resummation. 
These functions are regular at any finite value of $x_1\equiv k\mdot a$ and hence entire functions. \\[5pt]
As examples, we present the three-point amplitude from the Polyakov action, evaluated on higher modes $(2,1),  (2,2)$:
\begin{align}
\!\! A^{(2,1)}_{P}&=-i \black \kappa \, \Bigg[m^2 (v\mdot \veps )^2 \Bigg(\frac{5x_1^2}{72}  \, _1F_2\Big(\frac{3}{2};\frac{11}{4},\frac{13}{4};\frac{9 a_{t(2,1)}^2 x_1^2}{16 | a| ^2}\Big)\nn\\
   &~~~~+\frac{5}{63}  x_1^2 \, _2F_3\Big(\frac{3}{2},2;1,\frac{11}{4},\frac{13}{4};\frac{9 a_{t(2,1)}^2 x_1^2}{16 | a| ^2}\Big)\\
   &~~~~~~~~~+ \, _1F_2\Big(\frac{1}{2};\frac{3}{4},\frac{5}{4};\frac{9 a_{t(2,1)}^2 x_1^2}{16 | a| ^2}\Big)\Bigg)\nn\\
   &~~~~~~~~+i m v\mdot \veps\,   k\mdot S\mdot \veps  \, _1F_2\Big(\frac{3}{2};\frac{7}{4},\frac{9}{4};\frac{9 a_{t(2,1)}^2 x_1^2}{16 | a| ^2}\Big)\Bigg]\nn\\
\!\! A^{(2,2)}_{P}&=  -i\black \kappa\, \Bigg[m^2 \left(v\mdot \veps\right){}^2 \Bigg(\frac{5x_1^2 }{84} \, _1F_2\Big(\frac{5}{2};\frac{11}{4},\frac{13}{4};\frac{9 a_{t(2,2)}^2 x_1^2}{4 | a| ^2}\Big)\nn\\
&~~~~~~~~~~~~~~~~~~+ \, _1F_2\Big(\frac{1}{2};\frac{3}{4},\frac{5}{4};\frac{9 a_{t(2,2)}^2 x_1^2}{4 | a| ^2}\Big)\Bigg)\\
&~~~~~~~~+i m v\mdot \veps \, k\mdot S\mdot \veps \, _1F_2\Big(\frac{3}{2};\frac{7}{4},\frac{9}{4};\frac{9 a_{t(2,2)}^2 x_1^2}{4 | a| ^2}\Big)\Bigg].\nn
    \end{align}
Here  $a_{t(2,1)}, a_{t(2,2)}$ are the radius of the disc for the modes $(2,1), (2,2)$ respectively. For other higher modes with $l-j>1$, we have not be able to derive an analytical resummed form of amplitude, but numerical resummations converge for a wide range of parameter values. \\[5pt]
From the three-point amplitude above, we also provide the first post-Minkowskian bending angle for a test particle in the aligned-spin configuration. We follow the conventions in \cite{Brandhuber:2021eyq, Brandhuber:2023hhl, Kosower:2018adc}; see Fig. 1. The formula for the bending angle is \\[-20pt]
\begin{figure}[hbt]
  \centering
  \begin{minipage}{0.4\linewidth} 
    \centering
    \begin{tikzpicture}[baseline={([yshift=-2.8ex]current bounding box.center)}]
    \tikzstyle{every node}=[font=\small]	
    \begin{feynman}
         \vertex (a) {\(~~v_1~~\)};
         \vertex [right=1.5cm of a] (f2) [dot]{};
         \vertex [right=1.cm of f2] (c){};
         \vertex [above=1.5cm of a](ac){$~v_2,a~$};
         \vertex [right=1.5cm of ac] (ad) [dot]{};
         \vertex [above=1.5cm of c](cc){};
         \vertex [above=0.75cm of a] (cutL0);
         \vertex [right=0.99cm of cutL0] (cutL);
         \vertex [right=0.8cm of cutL] (cutR);
         \diagram* {
         (a) -- [fermion,thick] (f2)-- [fermion,thick] (c),
         (ad)--[photon,ultra thick,momentum=\(q\)](f2), 
         (ac) -- [fermion,thick] (ad)-- [fermion,thick] (cc), 
         (cutL)--[dashed, red,thick] (cutR)
         };
    \end{feynman}  
    \end{tikzpicture}
  \end{minipage}
  \begin{minipage}{0.53\linewidth} 
    \centering
    \begin{tikzpicture}
      \node at (5.,0.) {\includegraphics[width=0.83\linewidth]{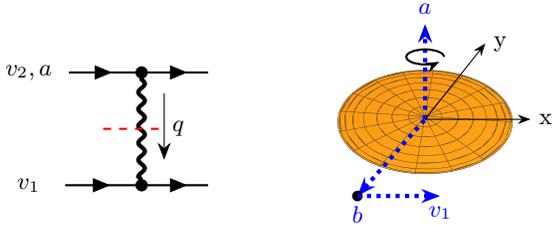}};
      \draw[dotted, ultra thick,blue,{Stealth[scale=.7]}-]  (4.2,-0.9)node[anchor=north] {$b$}-- (5.1,0.12) ;
      \fill[black] (4.2,-0.9) circle (2pt);
      \draw[ dotted, ultra thick,blue,-{Stealth[scale=.7]}]  (4.2,-0.9)-- (5.3,-0.9)node[anchor=north] {$v_1$} ;
      \draw[dotted, ultra thick,blue,-{Stealth[scale=.7]}] (5.1,0.12) -- (5.1,1.39) node[anchor=south] {$a$};
      \draw[-{Stealth[scale=1.]}] (5.1,0.12) -- (6.5,0.12) node[anchor=west] {\text{x}};
      \draw[-{Stealth[scale=1.]}] (5.1,0.12) -- (5.9,1.13) node[anchor=west] {\text{y}};
      \draw[thick,-{Stealth[scale=.8]},pos=0.1] (5.35,0.95) arc[start angle=0,end angle=320,x radius =0.25, y radius =0.1] ;
    \end{tikzpicture}
  \end{minipage}\\[-35pt]
  \caption{The related vectors are $v_{2}^\mu=(1,0,0,0)$, $v_{1}^\mu=(y,\sqrt{y^2-1},0,0)$, $b^\mu=(0,0,-|b|,0)$, $a^\mu=(0,0,0,|a|)$\,. }\label{fig:2}
\end{figure}
\begin{align}
	&\chi^{(l,j)} =-{\sqrt{s}\over 4m_1m_2\sqrt{y^2-1}}\times  \\
	&{\partial\over \partial|b|}\Big({1\over 4m_1m_2\sqrt{y^2-1}}\int {d^{D-2}q\over (2\pi)^{D-2} }e^{-i q\mdot b} (-i\mathcal{M}_{\rm HEFT}^{(l,j)}(q))\nn\Big)
\end{align}\\[-5pt]
where the impact parameter $b=b_1-b_2$ and 
\begin{align}
\begin{split}
\mathcal{M}^{(l,j)}_{\rm HEFT}(q)={i\over q^2}\!\!\!\!\!\!\!\!\sum_{\substack{\text{on-shell} \\ \text{graviton states}}} \!\!\!\!\!\!\!A_{3}^{(l,j)}(v_2,a,q)A_3(v_1,-q)
\end{split}\,.
\end{align}
Here $A_3(v_1,-q)=-i\kappa(m_1 v_1\mdot \veps)^2$ denotes the three-point gravity amplitude with a spinless particle. The complete first post-Minkowskian bending angle computed from the Polyakov action with the $(2,1)$-mode reads 
\begin{align}
&\chi^{(2,1)}_P = \frac{\kappa ^2 \sqrt{s}}{16 \pi ^2 \left(y^2-1\right)}\pi \Bigg[ \nn\\
&\left(2 y^2 - 1\right) \Bigg( \,_3F_2\Big(3,\frac{7}{2},\frac{7}{2};\frac{19}{4},\frac{21}{4};\frac{9 a_{t(2,1)}^2}{4 | b| ^2}\Big) \frac{2700   | a| ^2 a_{t(2,1)}^4}{17017 | b| ^7} \nn\\
&+ \,_3F_2\Big(1,\frac{3}{2},\frac{3}{2};\frac{7}{4},\frac{9}{4};\frac{9 a_{t(2,1)}^2}{4 | b| ^2}\Big) \frac{3  a_{t(2,1)}^2}{5 | b| ^3} \nn \\
&+ \,_3F_2\Big(2,\frac{5}{2},\frac{5}{2};\frac{15}{4},\frac{17}{4};\frac{9 a_{t(2,1)}^2}{4 | b| ^2}\Big)  \frac{1395  | a| ^2 a_{t(2,1)}^2}{4004 | b| ^5} \nn \\
&+ \,_3F_2\Big(1,\frac{3}{2},\frac{3}{2};\frac{11}{4},\frac{13}{4};\frac{9 a_{t(2,1)}^2}{4 | b| ^2}\Big)  \frac{25 | a| ^2}{84 | b| ^3} + \frac{1}{| b|} \Bigg)  \\
&- y\sqrt{y^2-1} \Bigg( \,_3F_2\Big(\frac{3}{2},2,\frac{5}{2};\frac{11}{4},\frac{13}{4};\frac{9 a_{t(2,1)}^2}{4 | b| ^2}\Big)  \frac{12 | a|  a_{t(2,1)}^2}{7 | b| ^4} \nn \\
&+ \,_3F_2\Big(\frac{1}{2},1,\frac{3}{2};\frac{7}{4},\frac{9}{4};\frac{9 a_{t(2,1)}^2}{4 | b| ^2}\Big)  \frac{2  | a|}{| b| ^2} \nn \Bigg) \Bigg]
\end{align}
For the $(2,2)$-mode, the first post-Minkowskian bending angle is 
\begin{align}
&\chi^{(2,2)}_P=\frac{\kappa ^2 \sqrt{s}}{16 \pi ^2 \left(y^2-1\right)}\pi \Bigg[\nn\\
&\left(2 y^2-1\right) \Bigg(\frac{450  | a| ^2 a_{t(2,2)}^2 \, }{1001 | b| ^5} \, _3F_2\Big(2,\frac{5}{2},\frac{7}{2};\frac{15}{4},\frac{17}{4};\frac{9 a_{t(2,2)}^2}{| b| ^2}\Big)\nn\\
&+\frac{5   | a| ^2 }{42 | b| ^3}\, _3F_2\Big(1,\frac{3}{2},\frac{5}{2};\frac{11}{4},\frac{13}{4};\frac{9 a_{t(2,2)}^2}{| b| ^2}\Big)\nn\\
&+\frac{12   a_{t(2,2)}^2 \, }{5 | b| ^3}\, _3F_2\Big(1,\frac{3}{2},\frac{3}{2};\frac{7}{4},\frac{9}{4};\frac{9 a_{t(2,2)}^2}{| b| ^2}\Big)+\frac{1 }{| b| }\Bigg)\\
&-y\sqrt{y^2-1} \Bigg(\frac{4 | a|  \, }{| b| ^2}\, _3F_2\Big(\frac{1}{2},\frac{3}{2},2;\frac{7}{4},\frac{9}{4};\frac{9 a_{t(2,2)}^2}{| b| ^2}\Big)\nn\\
&+\frac{5 | a|  \, }{6 a_{t(2,2)}^2}\, _3F_2\Big(-\frac{1}{2},\frac{1}{2},1;\frac{3}{4},\frac{5}{4};\frac{9 a_{t(2,2)}^2}{| b| ^2}\Big)-\frac{5 | a| }{6 a_{t(2,2)}^2}\nn\Bigg)\Bigg]\, .
\end{align}\\[-5pt]
We note that the hypergeometric functions in the bending angle above are all of type $\,_pF_q$ with $p=q+1$, which are not entire functions. Moreover, the bending angles are divergent at $|b|=\frac{3}{2}a_{t(2,1)}$ and $|b|=3a_{t(2,2)}$, which indicates that the singular rings lie outside the boundary of the disc for higher modes; see below for illustrations. 
\begin{equation}
\begin{array}{|c|c|c|c|}\hline
    \text{mode} & \text{ring radius} & \text{disc radius} & \text{graph} \\ \hline
    (1,1) & |a| & |a| & 
    \begin{tikzpicture}[baseline=-0.5ex]
        \node at (0,0) {\includegraphics[width=0.1\linewidth]{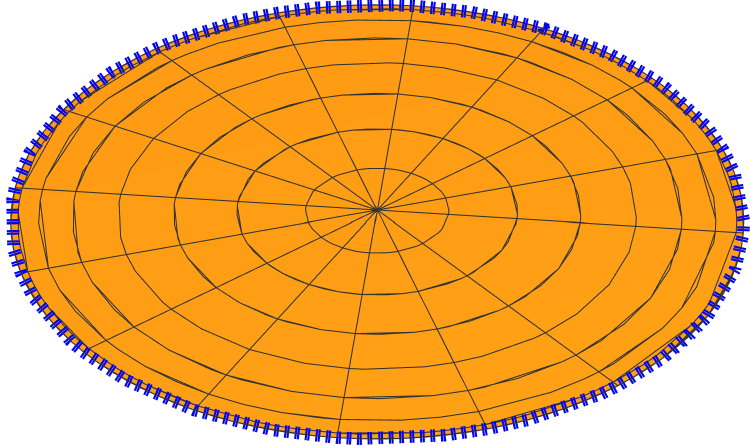}};  
        \draw[blue, thick] (0, 0) ellipse (0.42 and 0.23);
    \end{tikzpicture} \\  \hline
    (2,1) & \frac{3}{2}a_{t(2,1)} & a_{t(2,1)} & 
    \begin{tikzpicture}[baseline=-0.5ex]
        \node at (0,0) {\includegraphics[width=0.1\linewidth]{KerrY11.pdf}};
        \draw[blue, thick] (0, 0) ellipse (0.75 and 0.4);    
    \end{tikzpicture} \\ \hline
     (2,2) & 3a_{t(2,2)} & a_{t(2,2)} & 
    \begin{tikzpicture}[baseline=-0.5ex]
        \node at (0,0) {\includegraphics[width=0.1\linewidth]{KerrY11.pdf}};
        \draw[blue, thick] (0, 0) ellipse (1.0 and 0.6);    
    \end{tikzpicture} \\ \hline
\end{array}\, 
\end{equation}
\end{document}